\documentclass[sigconf]{acmart}
\usepackage{comment}
\usepackage{amsmath,amsfonts}
\usepackage{graphicx}
\usepackage{textcomp}
\usepackage{subcaption}
\usepackage{multirow}
\usepackage[linesnumbered,ruled,vlined]{algorithm2e}
\usepackage{xcolor}
\AtBeginDocument{%
  \providecommand\BibTeX{{%
    \normalfont B\kern-0.5em{\scshape i\kern-0.25em b}\kern-0.8em\TeX}}}

\begin{document}

\title{Evaluation of Posits for Spectral Analysis Using a Software-Defined Dataflow Architecture}

\author{Sameer Deshmukh}
\affiliation{%
  \institution{Tokyo Institute of Technology}
  \institution{NextSilicon Pvt. Ltd.}
  \city{Tokyo}
  \country{Japan}
}
\email{sameer.deshmukh@rio.gsic.titech.ac.jp}

\author{Daniel Khankin}
\affiliation{%
  \institution{NextSilicon Pvt. Ltd.}
  \city{Tel Aviv}
  \country{Israel}}
\email{daniel.khankin@nextsilicon.com}

\author{William Killian}
\affiliation{%
  \institution{NextSilicon Pvt. Ltd.}
  \city{Tel Aviv}
  \country{Israel}}
\email{will.killian@nextsilicon.com}

\author{John Gustafson}
\affiliation{%
  \institution{Arizona State University}
  \state{Arizona}
  \country{United States of America}}
\email{jlgusta6@asu.edu}

\author{Elad Raz}
\affiliation{%
  \institution{NextSilicon Pvt. Ltd.}
  \city{Tel Aviv}
  \country{Israel}}
\email{e@nextsilicon.com}

\begin{abstract}
Spectral analysis plays an important role in detection of damage in structures and deep learning. The choice of a floating-point format plays a crucial role in determining the accuracy and performance of spectral analysis. The IEEE Std 754\textsuperscript{TM} floating-point format (IEEE~754 for short) is supported by most major hardware vendors for ``normal'' floats. However, it has several limitations. The posit\textsuperscript{TM} format has been proposed as an alternative to IEEE~754. Previous work has attempted to evaluate posit format with respect to accuracy and performance. The accuracy of the posit has been established over IEEE~754 for a variety of applications. For example, our analysis of the Fast Fourier Transform shows 2x better accuracy when using a 32-bit posit vs. a 32-bit IEEE754 format. For spectral analysis, 32-bit posits are substantially more accurate than 32-bit IEEE~754 floats. Although posit has shown better accuracy than IEEE~754, a fair evaluation of posit with IEEE~754 format using a real hardware implementation has been lacking so far. A software simulation of posit format on an x86 CPU is about $\mathbf{69.3\times}$ slower than native IEEE~754 hardware for normal floats for a Fast Fourier Transform (FFT) of $\mathbf{2^{28}}$ points. We propose the use of a software-defined dataflow architecture to evaluate performance and accuracy of posits in spectral analysis. Our dataflow architecture uses reconfigurable logical elements that express algorithms using only integer operations. Our architecture does not have an FPU, and we express both IEEE~754 and posit arithmetic using the same integer operations within the hardware. On our dataflow architecture, the posit format is only $\mathbf{1.8\times}$ slower than IEEE~754 for a Fast Fourier Transform (FFT) of $\mathbf{2^{28}\approx 268}$ million points. This performance is achieved even though the number of operations for posit is almost $\mathbf{5\times}$ higher than IEEE~754. With this implementation, we empirically propose a new lower bound for the performance of posit compared to IEEE~754 format.
\end{abstract}

\maketitle

\section{Introduction}
\label{sec:introduction}

Spectral analysis is used in a wide variety of applications such as detection of damage in structures~\cite{Palacz2018} and deep learning~\cite{Bronstein_2017, bruna2014spectral}. The choice of format used for representation of real numbers with a given precision plays an important role in the performance and accuracy of spectral analysis. GNU MPFR~\cite{Fousse2007} can be used for extended precision calculations; however it is very inefficient for large problems. The IEEE~754 standard~\cite{noauthor_ieee_2019} is a legacy (1980s) floating-point format that has been adopted by major hardware vendors who have released hardware implementations using highly optimized circuits. However, the IEEE~754 format has several limitations such as non-associative addition and rounding error accumulation~\cite{goldberg1991}. It also uses a fixed number of bits for the fraction and exponent, which creates flat relative accuracy that is unrepresentative of the requirements of applications. 

The main computational kernel within spectral analysis is the Fast Fourier Transform (FFT)~\cite{trefethen2000}. Significant research has been dedicated to improving the accuracy of the FFT with IEEE~754 arithmetic. The FFT can be expressed as matrix multiplication operations and combined with techniques such as the Ozaki scheme~\cite{ozaki2012}, cascading GEMM~\cite{parikh2023} and mixed precision hardware~\cite{ootomo2023} to improve accuracy within the constraints of the IEEE~754 format. FFT implementations such as FFTX~\cite{franchetti2018} pre-compute the twiddle factors and reorder the computations to reduce rounding errors. 

Another area of research for improved accuracy of the FFT has been the use of a different implementation of floating point numbers (or \emph{floats}, for short) using the same bit-length as IEEE~754. Alternative implementations such as \emph{posits}~\cite{gustafson2022} have shown better accuracy for a wide variety of applications~\cite{montero2019, chien2020, murillo_comparing_2022}, including the FFT~\cite{ciocirlan2021}. Posits overcome some of the limitations of IEEE~754 with fully deterministic operations and more information-per-bit in the sense of higher Shannon entropy.

FFT speed is memory bandwidth bound. Although von Neumann memory access designs in CPUs and GPUs have been the mainstay of scientific computing, such architectures have proven to be inefficient for bandwidth-bound applications. Moreover, most of the power consumed by such architectures is used for loading and storing data from the caches and memory instead of performing calculations. FPGAs can overcome this memory access inefficiency with non-von Neumann memory access designs; however, they are hard to program.

Various hardware implementations of posits using RISC\nobreakdash-V~\cite{sharma2021} and FPGAs~\cite{carmichael2019} provide efficient hardware-level multiplication and addition routines for posits. The training of neural networks (NN) has been compared using floats and posits~\cite{montero2019, Silva2023}. The comparison by Montero et~al.~\cite{montero2019} uses VLSI synthesis to compare circuits for IEEE~754 and posit arithmetic; De~Silva et~al.~\cite{Silva2023} uses a Xilinx FPGA to compare NN accuracy. RISC-V implementations of posit hardware such as CLARINET~\cite{sharma2021} compare posits with IEEE~754 floats for various important algorithms on simulations of a RISC\nobreakdash-V CPU. The FFT using posits and IEEE~754 floats has previously been compared using a software-only implementation for small problem sizes~\cite{leong2023}.

Although previous attempts shed some light on the pros and cons of posits for various important applications, those attempts fall short on several fronts. The aforementioned RISC-V implementations do not run the algorithms on real silicon, and the FPGA implementations do not report a detailed performance analysis. The non-RISC-V software implementations pit their posit algorithms written in a high-level language (using only integer instructions) against IEEE~754 implementations (normal floats only) that have been optimized by floating-point unit (FPU) hardware designers over several decades. Thus, a fair comparison of the accuracy and performance between posit and IEEE~754 floats on real silicon for an important application has been lacking in the current literature.

In this paper, we compare performance and accuracy of 32-bit posit (\emph{posit32}) and normal IEEE~754 32-bit formats (\emph{float32}) using a new software-defined dataflow architecture. We test our implementation of these formats using the FFT, and then extend this to spectral analysis. Our architecture has no FPU. All operations are expressed using only a subset of basic integer arithmetic operations that are supported by hardware. With such an architecture we are able to tailor the hardware implementation to the required number format using software. Therefore, we are able to express both posit and normal IEEE~754 formats using basic arithmetic operations. This allows us to make use of our unique software-defined dataflow architecture to make an unbiased comparison between posit and IEEE~754 formats. 

Note that there is no dedicated hardware for arithmetic operations on posit32 or float32 in our hardware. Although the ideal comparison between the two formats would be the use of an optimized hardware implementation for both, this would be orthogonal to the objectives of this work. In this paper, we leverage our dataflow architecture to build hardware implementations using integer operations for both posit32 and float32 in order to make the first fair comparison between the posit32 and float32 formats.

Our architecture uses a non-von Neumann memory model and maps any algorithm to a directed acyclic graph (DAG). Each node of the graph can be a basic compute operation, a load/store operation, or a register. Unlike CPUs and GPUs, our architecture is much more efficient for memory-bound applications. Unlike FPGAs, our architecture is much easier to program as a result of leveraging compiler optimization of a high-level language like C for generation of the DAG. We make the following contributions in this paper:
\begin{enumerate}
    \item Analysis of the achievable accuracy when using posit32 and float32 for spectral analysis and the FFT. Our implementation of FFT with posit32 shows more than $2\times$ better accuracy compared to float32.
    \item A comparison of the performance and implementation cost of posit32 arithmetic compared to float32 (normal) arithmetic using basic integer operations by employing our software-defined dataflow architecture. We experimentally prove that the performance of the FFT using posits can almost match that of IEEE~754 (normals only) using our dataflow architecture, reaching within $\mathbf{1.82\times}$ of IEEE~754 for a problem size of $\mathbf{2^{28}\approx 268}$ \textbf{million}. 
\end{enumerate}

This is a large improvement over the difference in the CPU performance of $69.27\times$ for the same problem size between a software simulation of posit32 and a hardware implementation of float32. With our implementation, we empirically propose a new lower bound for the performance of posit compared to the IEEE~754 format.

\section{IEEE~754 Floating-Point Numbers}
\label{sec:ieee754-floating-point-numbers}

This section is specific to 32-bit (single-precision) normal floats, hereafter called float32. A normal IEEE~754 binary float is of the form

\begin{equation}
    x=(-1)^s \cdot m\cdot2^{e}
    \label{eq:normalfloat-repr}
\end{equation}

\noindent
where $s$ is the sign bit ($0$ or $1$), $m$ is the \emph{significand}\footnote{The significand is sometimes called the \emph{mantissa}, but the use of the term mantissa is discouraged and should be used in the context of logarithms.}  that satisfies $1 \le \left| m \right| < 2$, and $e~\in~\{e_\text{min}, \ldots, e_\text{max}\}$ is an integer that represents the exponent. There are $2^{25}=33554432$ exception cases to eq.~\ref{eq:normalfloat-repr}. for float32.

\begin{figure}
     \centering
     \includegraphics[width=\linewidth]{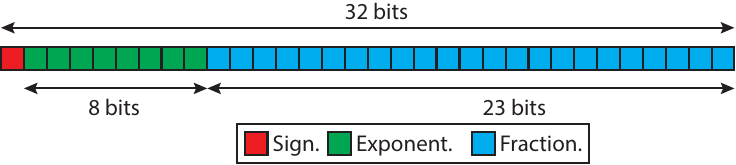}
     \caption{Bit sequence of a 32-bit IEEE~754 float. Unlike posits, IEEE floats use a fixed number of bits for the exponent and fraction.}
    \label{fig:ieee-754-repr}
\end{figure}

 The exponent $e$ is an 8-bit field interpreted as an unsigned integer biased by $127$ for normal floats, so $-126 \leq e \leq 127$. The minimal exponent value is reserved for the number $0$ and \textit{subnormal} numbers\footnote{formerly called \textit{denormalized} numbers}. The maximal exponent value is reserved for representing the special values \textit{infinity} and \textit{Not a Number} (NaN), the latter of which is used to represent values that are not real numbers (for example, $\sqrt{-1}$).

For normal numbers, $m \in [1,2)$. The leading bit (also called the \textit{implicit} or the \textit{hidden} bit) of the significand is always $1$ and is not explicitly stored. In the case of float32, the significand has $p=24$ bits. Subnormal numbers have the minimal exponent but the fraction part is different from zero. The implicit bit in this case is $0$ and the exponent is set to $e_\text{min} = -126$ so that we obtain

\begin{equation}
    x=(-1)^s \cdot m\cdot2^{-126}
    \label{eq:subnormalfloat-repr}
\end{equation}

\noindent
\textbf{This study considers only normal floats.} Almost all CPU and GPU hardware FPUs trap exception cases (subnormals, infinities, NaNs) and handle them with software or microcode, which run much slower. Hardware support for the entire IEEE 754 specification would require a massive increase in resources. The popular option is to disable exception trapping to increase execution speed.

\section{The Posit Specification}
\label{sec:the-posit-specification}

A posit is either not-a-real (NaR) or a real number $x$ of the form $2^M \times K$, where $K$ and $M$ are integers. Given $n$ total bits, the smallest positive posit value, \textit{minPos}, is $2^{-4n+8}$ and the largest positive posit value, \textit{maxPos}, is $1/\textit{minPos}=2^{4n-8}$. We here consider $n=32$, also called posit32. Unlike IEEE 754, there are no redundant representations; every posit maps to a unique real number or NaR. $0$ is represented by all bits to $0$, and NaR by only the most significant bit set to 1 followed by all $0$ bits. The general format for posit encoding is as follows:
\begin{itemize}
    \item $S$ - sign bit; represents an integer of value $0$ or $1$.
    \item $R$ - regime bit field; sequence of identical bits of length $k$, i.e., $R_{k-1} = \ldots = R_0$. The sequence is terminated either by $\bar{R_0}=1 - R_0$ or by the total length of bits. The field represents an integer $r$ of value $-k$ if $R_0$ is $0$ or $k-1$ if $R_0$ is $1$.
    \item $E$ - the exponent bit field has length 2 bits, representing an unsigned integer $e$ of value in $[0, 3]$. If one or both bits are pushed out by the regime field, those bits are $0$.
    \item $F$ - fraction bit field; has length of $\max(0, n-5)$ bits, representing an $m$-bit unsigned integer $M$ divided by $2^m$ resulting in a fraction value $f$ in $[0, 1)$.
\end{itemize}

\begin{figure}
    \centering
    \includegraphics[width=\linewidth]{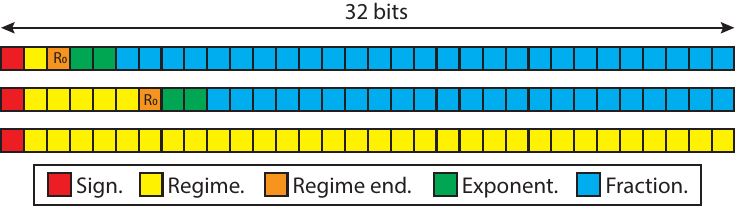}
    \caption{Posit32 with varying number of bits for the regime, exponent, and fraction. The uppermost diagram uses all the $27$ bits available for the fraction. The middle diagram has more bits for the regime, but fewer fraction bits. The lowermost diagram maximizes the use of the regime bits.}
    \label{fig:posit-specification}
\end{figure}

A posit is represented by the function shown in Eq.~\ref{eq:posit-repr}. The power $4r+e$ is the \textit{scaling factor} of the posit and the $f$ is the fraction. While IEEE 754 uses sign-magnitude, posits use 2's complement. All reals are normalized and have an implicit $1$, similar to normal IEEE 754 floats. Unlike IEEE 754 formats, posits have variable length bit fields that are determined by the value being stored. Fig.~\ref{fig:posit-specification} demonstrates three possible configurations for a posit32. Values with scaling factors near unity allow a greater number of bits for the significand than values with extreme exponents. This works well for most applications, including our target FFT application as will be shown in Sec.~\ref{sec:results}. For positive posits,
\begin{equation}
    x = (1.f) \times 2^{4r + e}
    \label{eq:posit-repr}
\end{equation}

Negative posits values can be decoded by negating the bits as a 2's complement integer, decoding as a positive posit, and negating the $x$ result.

Standard posits support an \emph{exact dot product} using a fixed-point format that is $16$ times as large as the posit precision. Our present implementation does not support this feature.

Posit32 is between $1$ and $16$ times as accurate as float32 for real values $x$ such that $2^{-20}\le |x| \le 2^{20}$, that is, magnitude of about one-millionth to one million. Outside that range, posit32 has lower accuracy than float32. The dynamic ranges are very similar for posit32 and float32.

\subsection{Posit algorithms}
\label{sec:posit-decode}

Like floats, posits~\cite{gustafson_beating_2017} require decoding before arithmetic operations. Alg.~\ref{alg:p32-decode} shows the procedure for decoding a posit into its sign, scaling factor, and fraction parts. Unlike IEEE~754, posits have only two special states: 0 (all bits set to $0$) and NaR (all bits except the sign bit set to 0). Hence, there is no need to trap exceptions as is done for IEEE~754.

Alg.~\ref{alg:p32-addition} shows the algorithm we use to add two posits. The decoding step defined in Alg.~\ref{alg:p32-decode} is used as the \texttt{decode} function in this algorithm. Note that we handle the special cases of $0$ and NaR separately in the addition. Subtraction is done by adding the 2's complement of the subtrahend. Alg.~\ref{alg:p32-multiplication} shows the algorithm for the product of two posits. The fraction and scaling factor generated is encoded back into a posit after rounding of the fraction~\cite{gustafson2022}.

\begin{algorithm}
    \SetAlgoLined
    \LinesNumbered
    \SetNoFillComment
    \DontPrintSemicolon
    \KwIn{\textit{posit}}
    \KwOut{\textit{sign}, \textit{sf}, \textit{fraction}, \textit{state}}
    \SetKwFunction{FDecode}{decode}
    \SetKwProg{Fn}{Function}{:}{end}{}
    \Fn{\FDecode{$\textit{posit}$}} {
        \If{$\textit{posit} = 0$} {
            $\textit{state} \gets 0$ \tcc{Found zero.} 
            \Return $\textit{sign}=0$, $\textit{exp}=0$, $\textit{fraction}=0$, \textit{state}
        } 
        \If{$\textit{posit} = (1 \ll 31)$} {
            $\textit{state} \gets -1$ \tcc{Found NaR.} 
            \Return $\textit{sign}=1$, $\textit{exp}=0$, $\textit{fraction}=0$, \textit{state}
        }
        \tcc{Save the sign of the posit.}
        $\textit{sign} \gets \textit{posit} \mathbin{\&} (1 \ll 31)$  \\
        $r \gets \textit{posit}  \mathbin{\&} (1 \ll 30)$ \tcc{Regime sign.}
        $\textit{temp} \gets \textit{posit} \ll 1$ \\
         \tcc{Count leading 1s or 0s.}
        \uIf{r} {
            $l \gets $ CLZ($\neg \textit{temp}$) \\
        }
        \Else {
            $l \gets $ CLZ(\textit{temp}) \\
        }
        $\textit{temp} \gets \textit{temp} \ll l+1$ \\
        \uIf{r} { \tcc{Regime magnitude.} 
            $k \gets -l$ 
        }
        \Else {
            $k \gets l-1$
        }
        $e \gets \textit{temp} \gg 30$ \tcc{Exponential.} 
        $\textit{fraction} \gets \textit{temp} \gg 2$ \tcc{Fraction.} 
        \tcc{Scaling factor.} 
        $\textit{sf} \gets 4 \times k + e$  \\
        \tcc{Non-zero and non-NaR.} 
        $\textit{state} \gets 1$  \\
        \Return \textit{sign}, \textit{exp}, \textit{fraction}, \textit{state}}
  \caption{Decoding algorithm for posit32. The \textit{state} is set to $0$ if the posit is $0$, $-1$ if NaR, and $1$ for all other values. The scaling factor is \textit{sf}.}
  \label{alg:p32-decode}
\end{algorithm}

\begin{algorithm}
    \SetAlgoLined
    \LinesNumbered
    \SetNoFillComment
    \DontPrintSemicolon
    \KwIn{$\textit{posit}_1$, $\textit{posit}_2$}
    \KwOut{\textit{sum}}
    \SetKwFunction{FSum}{add}
    \SetKwProg{Fn}{Function}{:}{end}{}
    \Fn{\FSum{$\textit{posit}_1$, $\textit{posit}_{2}$}} {
        $s_1, \textit{sf}_1, \textit{frac}_1, \textit{state}_1 \gets \texttt{decode}(\textit{posit}_1)$\;
        $s_2, \textit{sf}_2, \textit{frac}_2, \textit{state}_2 \gets \texttt{decode}(\textit{posit}_2)$\;
        \tcc{One number is zero.}
        \If{$\textit{state}_1 = 0$} {
            \Return $\textit{posit}_2$\;
        }
        \If{$\textit{state}_2 = 0$} {
            \Return $\textit{posit}_1$\;
        }
        \tcc{Output is NaR.}
        \If{$\textit{state}_1 = -1 \lor \textit{state}_2 = -1$} {
            \Return $1 \ll 31$\;
        }
        \tcc{Add implicit bit to each.}
        $\textit{frac}_1 = (1 \ll 31) \mathbin{|} (\textit{frac}_1 \gg 1)$\;
        $\textit{frac}_2 = (1 \ll 31) \mathbin{|} (\textit{frac}_2 \gg 1)$\;
        \tcc{Calculate fraction shift.}
        \uIf{$\textit{sf}_1 \le \textit{sf}_2$} {
            $\textit{shift} \gets \textit{sf}_2 - \textit{sf}_1$\;
            $\textit{frac}_2 = \textit{frac}_2 \gg \textit{shift}$\;
            $\textit{sf} = \textit{sf}_2$\;
        }
        \Else {
            $\textit{shift} \gets \textit{sf}_1 - \textit{sf}_2$\;
            $\textit{frac}_1 = \textit{frac}_1 \gg \textit{shift}$\;
            $\textit{sf}=\textit{sf}_1$\;
        }
        \tcc{Fraction addition}
        $\textit{frac}_{\textit{sum}}, \textit{carry} \gets \textit{frac}_1 + \textit{frac}_2$\;
        \uIf{$\textit{carry}$} {
            $\textit{sf} = \textit{sf} + 1$\;
        }
        \Else {
            $\textit{frac}_{\textit{sum}} = \textit{frac}_{\textit{sum}} \ll 1$\;
        }
        \Return $\texttt{encode}(s_1 \oplus s_2, \textit{sf}, \textit{frac}_{\textit{sum}})$\;
    }
    \caption{Addition algorithm for posit32 inputs.}
    \label{alg:p32-addition}
\end{algorithm}

\section{Dataflow Architecture}
\label{sec:dataflow-architecture}

\subsection{Comparison of a dataflow architecture with a CPU}
\label{sec:compare-data-flow-with-pipeline}

\begin{figure}
    \centering
    \includegraphics[width=\linewidth]{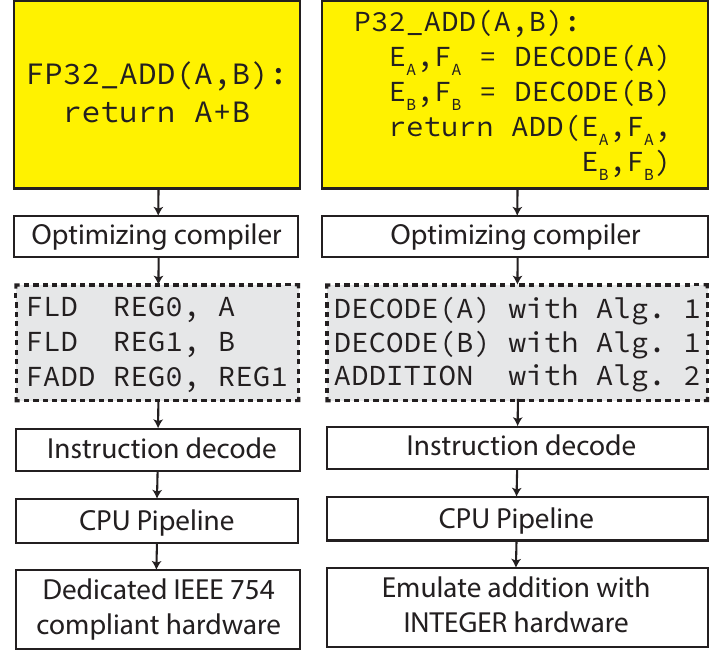}
    \caption{Comparison of the operations performed with floats on the left vs. posits on the right using a general-purpose CPU for addition. CPUs have dedicated hardware for normal IEEE~754 floats whereas posits are emulated on integer hardware.}
    \label{fig:cpu-add}
\end{figure}

Fig.~\ref{fig:cpu-add} illustrates a simple addition function on a general-purpose CPU. The left side shows the function \texttt{FP32\_ADD} that adds two float32 values, which can be translated into three CPU instructions ($2\times$\texttt{FLD} (load float) and another \texttt{FADD} (add float)) by an optimizing compiler. These instructions are executed by hardware specifically dedicated to IEEE~754 arithmetic for normal floats. On the other hand, the right side shows a function \texttt{P32\_ADD} that performs the addition of two posit32 values. Dedicated hardware for posit arithmetic does not exist on general-purpose CPUs. Therefore, posit arithmetic must proceed with dedicated algorithms such as \texttt{decode} and \texttt{add}. The optimizing compiler will then convert these functions into instructions that will be evaluated by hardware meant for integer calculations. Although Fig.~\ref{fig:cpu-add} is made by keeping in mind the CPU execution model, the same will apply to any device that makes use of a von Neumann memory model and uses dedicated instruction pipelines and execution units for floating point instructions, such as a GPU.

In contrast to the general-purpose CPU, we consider a dataflow architecture, where a program is represented by a directed acyclic graph (DAG). The nodes of a DAG represent operators that are applied to data objects. The edges of the DAG represent the movement of data objects between the operators. An operator can have several outputs and several or no inputs. The DAG is generated using a special compiler from a high-level language such as C. The DAG is then mapped by the managing software to an array of Logical Elements (LEs) on hardware. Initially, the LEs on hardware are generic and the managing software is responsible for reprogramming the generic LEs into target operators and activating the appropriate interconnections between them.  Broadly, the LEs can be either arithmetic/logical operations such as ADD or SELECT, or they can be registers or memory operations.

\begin{figure}
    \centering
    \includegraphics[width=\linewidth]{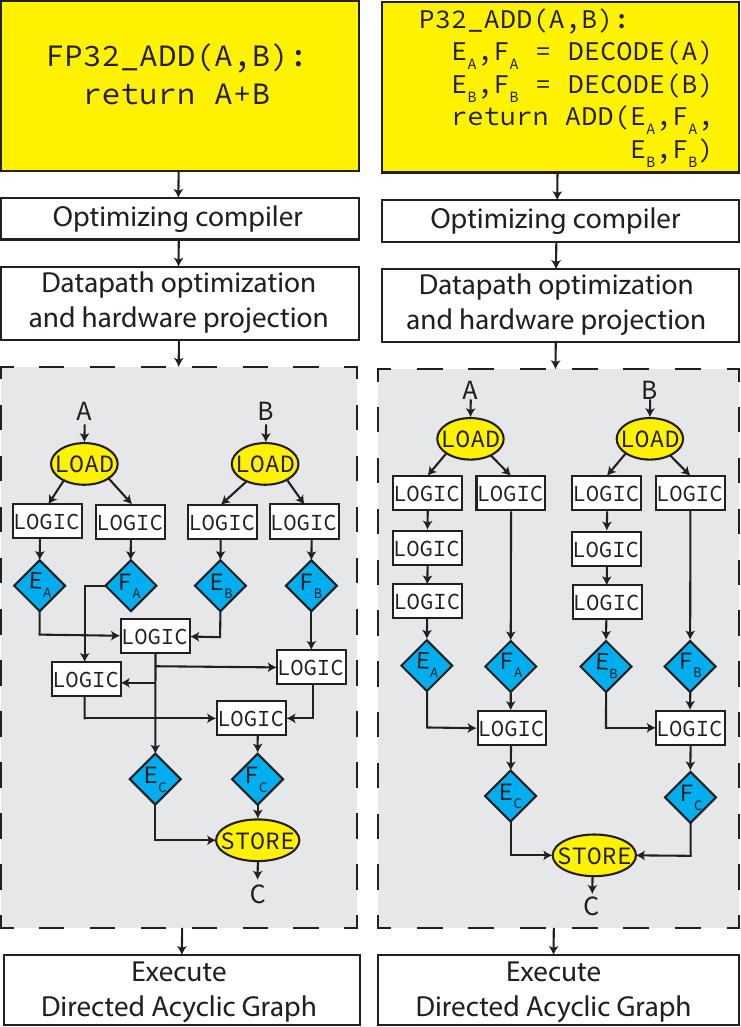}
    \caption{Comparison of addition of float32 on the left and posit32 on the right on a software-controlled dataflow architecture. The computation is projected onto the hardware using an optimizing compiler and hardware projection software stack, and then executed as per the flow of data.}
    \label{fig:dataflow-add}
\end{figure}

Our hardware only supports a subset of integer arithmetic operations. The set of operators may seem restrictive. However, the main purpose of using such software-defined architecture is the decoupling of the operational and computational complexity from the underlying hardware implementation. In other words, similar to other software-defined concepts, the goal is to use generic hardware elements and offload the computational complexity to software.

Fig.~\ref{fig:dataflow-add} shows an illustration of addition on a dataflow architecture for floats on the left and posits on the right. Contrary to Fig.~\ref{fig:cpu-add}, the optimizing compiler passes into a datapath optimization and hardware projection layer that generates the DAG shown in the dotted gray box. The nodes of this graph represent the operations being performed by the hardware and the edges are the data dependencies between them. Nodes that do not have dependencies can execute in parallel whereas those with dependencies must execute in serial.

The off-white rectangular nodes represent arithmetic and logical operations. The blue rhombus-shaped nodes represent registers, and the yellow ovals are memory operations. In contrast to  the CPU implementation shown in Fig.~\ref{fig:cpu-add}, both float and posit implementations are using the nodes of the dataflow architecture. Although IEEE~754 exceptions are ignored here, the implementation allows us to compare the two number formats and test them with various algorithms, as will be further  discussed in Sec.~\ref{sec:arithmetic-operations-ieee754-posit}.

\begin{algorithm}
    \SetAlgoLined
    \LinesNumbered
    \SetNoFillComment
    \DontPrintSemicolon
    \KwIn{$\textit{posit}_1$, $\textit{posit}_2$}
    \KwOut{$product$}
    \SetKwFunction{FProduct}{\textit{multiply}}
    \SetKwProg{Fn}{Function}{:}{end} {}
    \Fn{\FProduct{$\textit{posit}_1$, $\textit{posit}_2$}} {
        $s_1, \textit{sf}_1, \textit{frac}_1, state_1 \gets$ \texttt{decode}($\textit{posit}_1$)\;
        $s_2, \textit{sf}_2, \textit{frac}_2, state_2 \gets$ \texttt{decode}($\textit{posit}_2$)\;
        \tcc{Test for NaR}
        \If {$state_1 = 1 \ll 31 \lor state_2 = 1 \ll 31$} {
            \Return NaR\;
        }
        \tcc{Test for zero.}
        \If {$state_1 = 0 \lor state_2 = 0$} {
            \Return 0\;
        }
        $\textit{sf} \gets \textit{sf}_1 + \textit{sf}_2$\;
        \tcc{XOR sign bits.}
        $\textit{sign} \gets s_1 \oplus s_2$\;
        \tcc{Add implicit bit.}
        $\textit{frac}_1 \gets (1 \ll 31) \mathbin{|} (\textit{frac}_1 \gg 1)$\;
        $\textit{frac}_2 \gets (1 \ll 31) \mathbin{|} (\textit{frac}_2 \gg 1)$\;
        $\textit{frac} \gets \textit{frac}_1 \times \textit{frac}_2$\;
        \Return \texttt{encode}($\textit{sign}$, \textit{sf}, $\textit{frac},1$)\;
    }
  \caption{32-bit posit multiplication algorithm.}
  \label{alg:p32-multiplication}
\end{algorithm}

\subsection{Expression of a DAG on our dataflow architecture}
\label{sec:mapping-a-dag-to-hardware}

Fig. \ref{fig:architecture-overview} shows an overview of the layout of our software-defined dataflow architecture. The whole chip is vertically split into \emph{tiles}, each of which is divided into 32 \emph{clusters}, which are shown in deep yellow. Each cluster contains several \emph{Logical Elements} (LEs) that can be programmed by software depending on the algorithm in use. The software adds edges between the LEs as the per the dependencies within the algorithm. The zoomed-in picture of the cluster on the right shows the interconnections between LEs on a single cluster. The edges can also spawn between LEs present on different clusters, therefore allowing a DAG to spread across multiple clusters. In addition to that, LEs between different tiles can communicate with each other using inter-tile edges, therefore allowing large DAGs to be accommodated on the chip using multiple tiles. All the tiles are connected to volatile memory shown in orange color on the left. The memory is accessed by special LEs programmed for memory operations (such as those shown in the oval shapes in Fig. \ref{fig:dataflow-add}.

As shown in Sec.~\ref{sec:compare-data-flow-with-pipeline}, an optimizing compiler can compile a program written in a high-level language such as C, and then project it onto our hardware as shown in Fig.~\ref{fig:architecture-overview}. This dataflow graph can then be executed with data loaded from the volatile memory. The execution of the program happens from the top of the chip to the bottom. Loops within a program go over the same graph as long as they need to iterate. Unlike a general-purpose CPU, dataflows within the chip use specialized LEs that act like registers. This non-von Neumann memory model is further elaborated in Sec.~\ref{sec:non-von-neumann-mem-access}.


\subsection{Memory access design}
\label{sec:non-von-neumann-mem-access}

General-purpose CPUs make use of the von Neumann memory model, where computing is separated from data storage such as cache and memory. A limited number of registers within each core are used for storage of intermediate results. The von Neumann design has found widespread adoption; however, it is extremely inefficient with regards to memory access, and does not perform well for memory-bound applications. Moreover, computer architectures that make use of von~Neumann designs typically spend a majority of their power consumption on fetching and storing data.

Our software-defined dataflow architecture uses a non-von~Neumann memory model. Intermediate results are stored within the chip by programming LEs to behave like registers, thereby saving a lot of cycles and energy that would have otherwise gone into fetching and storing data from memory. Data can be loaded and stored in these registers from volatile memory as shown in Fig.~\ref{fig:architecture-overview}. However, far fewer trips are needed between registers and memory in a non-von Neumann architecture.

General-purpose CPUs typically make use of uniform memory access, i.e., the latency for fetching a particular byte from anywhere in memory at a particular hierarchy is the same. However, our dataflow architecture has non-uniform memory access. This means that if a byte is stored in the north side of the volatile memory and an LE in the south side requests that data, it will accrue longer latency than if the byte was present on the south side of the volatile memory. This has implications for the achievable throughput of a DAG, as will be further discussed in Sec.~\ref{sec:performance-data-flow-architecture}.

\subsection{Threading model}
\label{sec:threading-model}

\begin{figure}
    \centering
    \includegraphics[width=\linewidth]{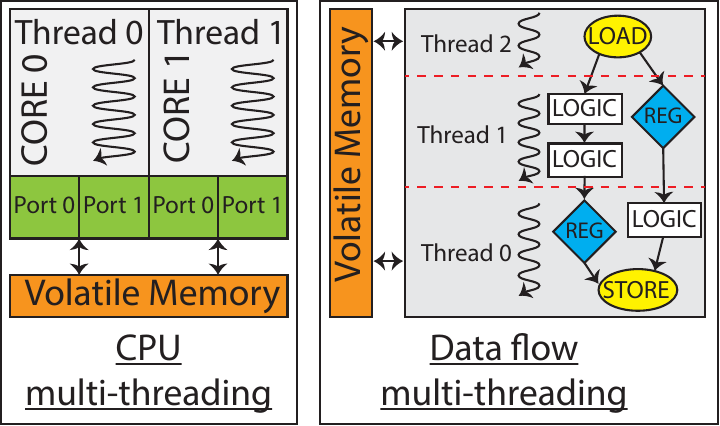}
    \caption{Multi-threaded execution on a general-purpose CPU (left) vs. our software-defined dataflow architecture (right).}
    \label{fig:threading-model}
\end{figure}

Fig.~\ref{fig:threading-model} shows the threading model of a general-purpose CPU on the left, and that of our dataflow architecture on the right. The CPU model on the left (assuming no Simultaneous Multi-Threading) runs a single thread per physical core of the CPU. Each core contains various `ports' that contain execution units such as the FPU or the load/store unit. The cores are then connected to a volatile memory. Each thread can execute independently. The throughput of a set of instructions depends on the speed of the execution units and the memory latency. All threads in the CPU can start and stop independently.

The dataflow architecture on the right of Fig.~\ref{fig:threading-model} has a different threading model. The execution of the DAG has to flow from top to the bottom, and the multi-threading in this case begins with thread 0, which first starts at the \texttt{LOAD} node at the top of the DAG, and works its way to the bottom. Thread~1 can start execution once thread 0 has moved down the DAG. Multiple threads can execute the same DAG in this manner. Each thread runs the same algorithm with different data that can be read from the volatile memory. This allows the dataflow architecture to execute parallel loops using the same DAG. Each thread has access to multiple execution units (similar to ports in a CPU) and can therefore execute multiple independent LEs in parallel. 

\begin{figure}
    \centering
    \includegraphics[width=\linewidth]{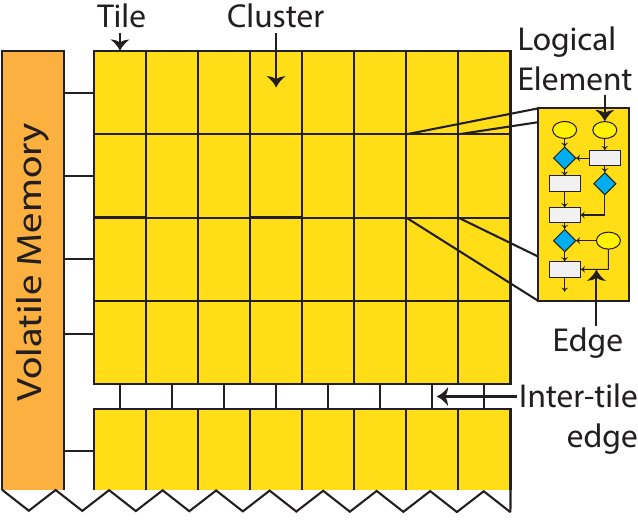}
    \caption{Block diagram of our software-defined dataflow architecture. The chip is vertically split into multiple tiles, each of which is divided into clusters. The cluster contains Logical Elements (LEs). The LEs can access data from volatile memory shown in orange on the left. Algorithms can spawn multiple clusters and tiles. The zoomed-in cluster on the right shows the connections between the software-programmed LEs.}
    \label{fig:architecture-overview}
\end{figure}

Assuming each LE takes one clock cycle to execute, the threading model of the dataflow architecture can ensure an ideal throughput of one clock cycle for any DAG. However, several reasons can reduce the throughput. For example, lack of data reuse between successive threads can lead to thread stalling as a result of excessive reads and writes from volatile memory. If a single thread stalls, the threads before must wait for it to resume. Tall DAGs that spawn multiple tiles (Fig. \ref{fig:architecture-overview}) can also lead to loss of performance because threads must save their context in one tile and reestablish it in the next, thus leading to further stalling. The number of execution units for a particular kind of LE is limited per thread. Therefore, excessive port pressure owing to many parallel LEs of the same kind can also contribute to reduced performance.

\section{Results}
\label{sec:results}

We first analyse the accuracy for the FFT and spectral method when using posit32 vs. float32 in Sec.~\ref{sec:analysis-of-accuracy-p32-f32}. We then build on this implementation and evaluate the use of float32 and posit32 arithmetic on our software-defined dataflow architecture. All operations for all number formats and algorithms are reported assuming that denormals, NaNs and infinity never occur in the computation (also known as `fast-math' mode in some compilers). Therefore, the parts of the algorithms for posit shown in Sec. \ref{sec:arithmetic-operations-ieee754-posit} that deal with NaR values are disabled when reporting results. All the code for our dataflow architecture is written in C, and compiled with the NextSilicon compiler, which converts the code into a DAG that can be projected on the chip. All the code for the CPU results is written in C and compiled using Clang 16.0 with compile options \texttt{-Ofast -march=native -fopenmp -lm}.

\subsection{Analysis of accuracy of posit32 vs. float32}
\label{sec:analysis-of-accuracy-p32-f32}

\subsubsection{Accuracy analysis of the FFT}
\label{sec:accuracy-analysis-fft}

\begin{figure}
    \centering
    \includegraphics[width=\linewidth]{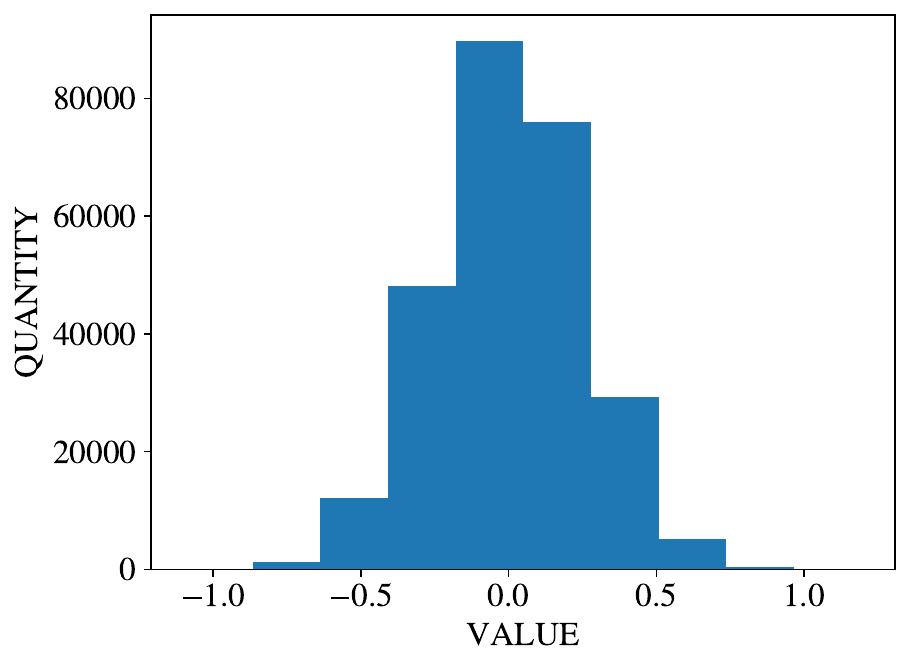}
    \caption{Histogram of the inputs of the FFT.}
    \label{fig:hist-inputs}
\end{figure}

Our posit implementation is compared with floats using an FFT implemented using the radix-4 iterative Stockham algorithm. The same algorithm is used for implementation of an Inverse Fast Fourier Transform (IFFT). We run the input through a 1D FFT followed by an IFFT in order to check the accuracy of the floating point representation scheme. The inputs are normally distributed random values within $[-1,1]$. Fig.~\ref{fig:hist-inputs} shows a histogram of the inputs to the 1D FFT for a vector size of $2^{18}$. The values are between the interval $[-1,1]$, where posits have the greatest accuracy advantage compared to IEEE~754~\cite{murillo_comparing_2022}.

\begin{equation}
    {\rm error} = \sqrt{\sum_{i=0}^{N}{(x_i - y_i)^2}}
    \label{eq:fft-error}
\end{equation}

\begin{figure}
    \centering
    \includegraphics[width=\linewidth]{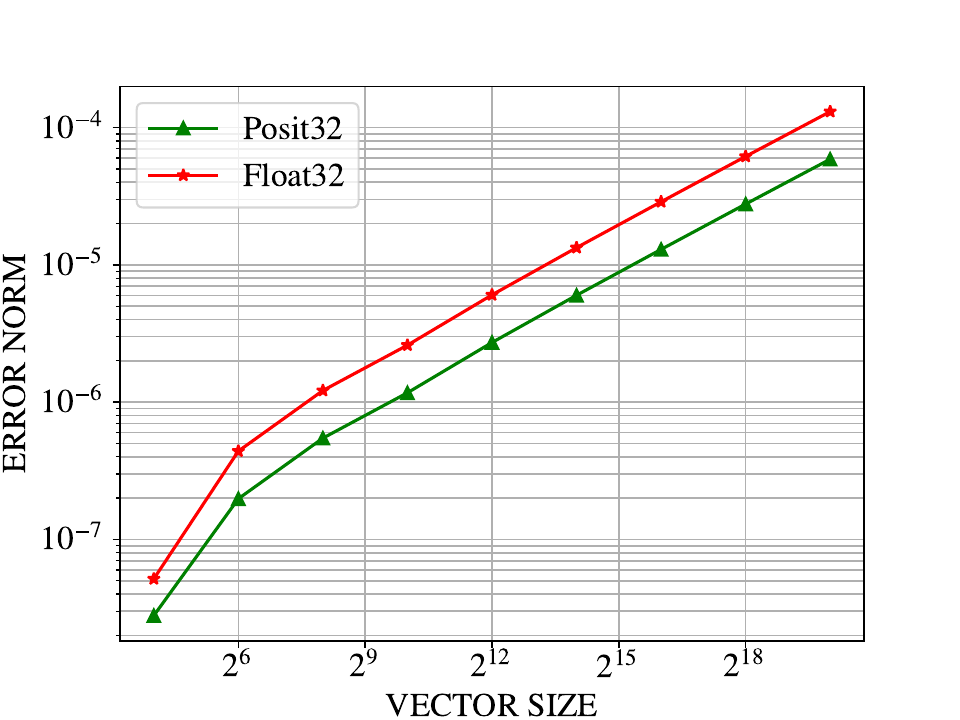}
    \caption{Accuracy of the FFT using posit32 vs. float32.}
    \label{fig:accuracy-evaluation}
\end{figure}

Fig.~\ref{fig:accuracy-evaluation} shows the comparison of accuracy between posit32 and float32. The accuracy of the FFT is calculated by passing a vector $x_N$ of length $N$ through a radix-4 FFT and IFFT as described in Sec.~\ref{sec:results} and obtaining an output vector $y_N$. The error is then calculated as shown in Eq.~\ref{eq:fft-error}. It can be seen that posit32 is consistently much more accurate than float32 for all vector sizes.

\subsubsection{Accuracy analysis of the spectral method}
\label{sec:spectral-method-analysis}

Spectral analysis can make use of the FFT for polynomial convolution~\cite{trefethen2000}, making the accuracy and performance of the FFT important for spectral analysis. We model a 1D wave in an isotropic medium using the spectral method and test it with our posit and float implementations. We assume a simple 1D Laplace kernel as our governing PDE. The input vector is denoted by $x$, where $x$ varies uniformly from $0$ to $N$ with a uniform interval of $\frac{2\pi}{N \cdot d}$,  where $d$ is a constant ($20$ in this case). We use $1000$ time steps for all experiments. The source wavelets for these experiments consist of sines and cosines, therefore guaranteeing convergence of the Fourier series.

\begin{figure}
    \centering
    \includegraphics[width=\linewidth]{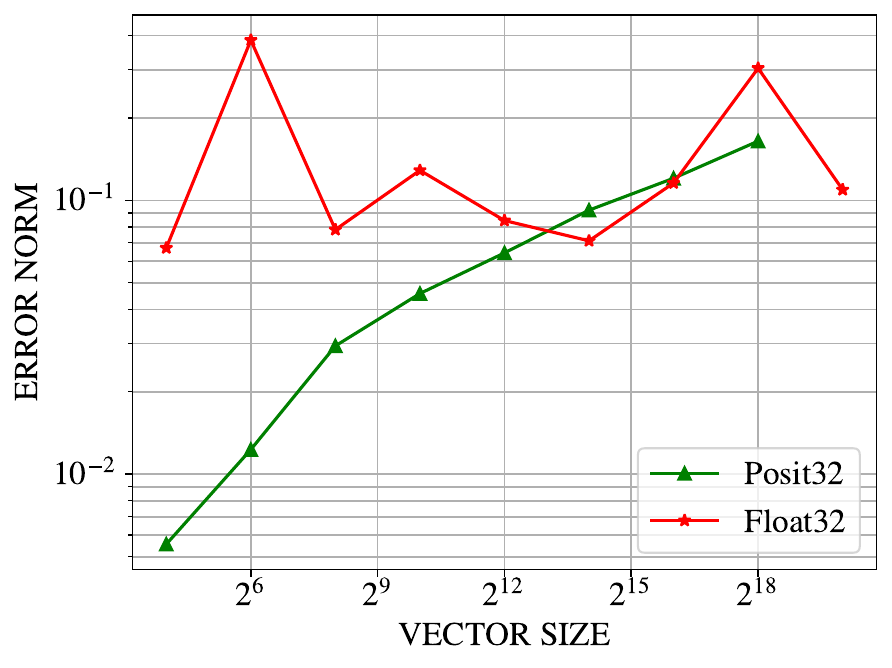}
    \caption{The error shown by a 1D spectral method when using posit32 and float32.}
    \label{fig:spectral-method-error-norm}
\end{figure}

Fig.~\ref{fig:spectral-method-error-norm} shows the error norm between the result of a spectral method calculation using full-precision numbers with the GNU MPFR library compared with posit32 and float32 implementations. The error norm on the the $y$-axis is calculated by running the spectral method using a $250$-bit precision using the MPFR library and then comparing it against the given floating point format using Eq.~\ref{eq:fft-error}.

The error norm of posit32 is better than that of the float32 for most cases. The values fed into the FFT are not necessarily close to \texttt{[-1,1]} as shown in Fig.~\ref{fig:hist-inputs}. Since the FFT and IFFT are the most intensive kernels within spectral analysis, we expect the performance on our dataflow architecture to be similar to that shown in Sec.~\ref{sec:evaluation-of-fft}.

\subsection{Evaluation of posit32 and float32 on our dataflow architecture}
\label{sec:evaluation-of-fft}

\begin{table*}[]
\centering. 
\begin{tabular}{|c|ccc|ccc|}
\hline
\multirow{2}{*}{\textbf{LE Description}} & \multicolumn{3}{c|}{\textbf{Posit32}}& \multicolumn{3}{c|}{\textbf{Float32}} \\ \cline{2-7} 
    & \multicolumn{1}{c|}{\textbf{ADD}} & \multicolumn{1}{c|}{\textbf{SUB}} & \textbf{MUL} & \multicolumn{1}{c|}{\textbf{ADD}} & \multicolumn{1}{c|}{\textbf{SUB}} & \textbf{MUL}\\ \hline
Min/max  & \multicolumn{1}{c|}{13}  & \multicolumn{1}{c|}{13}  & 12    & \multicolumn{1}{c|}{5}  & \multicolumn{1}{c|}{5}   & 2  \\ \hline
Integer arithmetic  & \multicolumn{1}{c|}{85}   & \multicolumn{1}{c|}{85}  & 66   & \multicolumn{1}{c|}{14}   & \multicolumn{1}{c|}{14}  & 5 \\ \hline
Bitwise operations & \multicolumn{1}{c|}{150}  & \multicolumn{1}{c|}{146}     & 104           & \multicolumn{1}{c|}{16} & \multicolumn{1}{c|}{17}    & 10 \\ \hline
Special proprietary LEs   & \multicolumn{1}{c|}{64} & \multicolumn{1}{c|}{66} & 48  & \multicolumn{1}{c|}{13}   & \multicolumn{1}{c|}{13}   & 5 \\ \hline
\textbf{Total} & \multicolumn{1}{c|}{\textbf{333}}   & \multicolumn{1}{c|}{\textbf{331}} &  \textbf{241} & \multicolumn{1}{c|}{\textbf{47}}   & \multicolumn{1}{c|}{\textbf{48}} &    \textbf{22} \\ \hline
\end{tabular}
\vspace{1mm}
\caption{Type of LEs used for the posit32 and float32 operators.}
\label{tab:le-distribution-basic-operators}
\end{table*}

Sec.~\ref{sec:analysis-of-accuracy-p32-f32} empirically proves that  posit is more accurate than IEEE~754 for the same bit-length for both the FFT and spectral method. CPU implementations of posit32 do not implement posit32 arithmetic in hardware, and are therefore much slower than float32. In this section, we will empirically prove that our posit32 implementation using dataflow hardware is much faster than the CPU implementation and then elaborate on the reasons of this improved performance.

\subsubsection{Performance of the FFT}
\label{sec:performance-data-flow-architecture}

\begin{figure}
    \centering
    \begin{subfigure}[b]{\linewidth}
        \includegraphics[width=\linewidth]{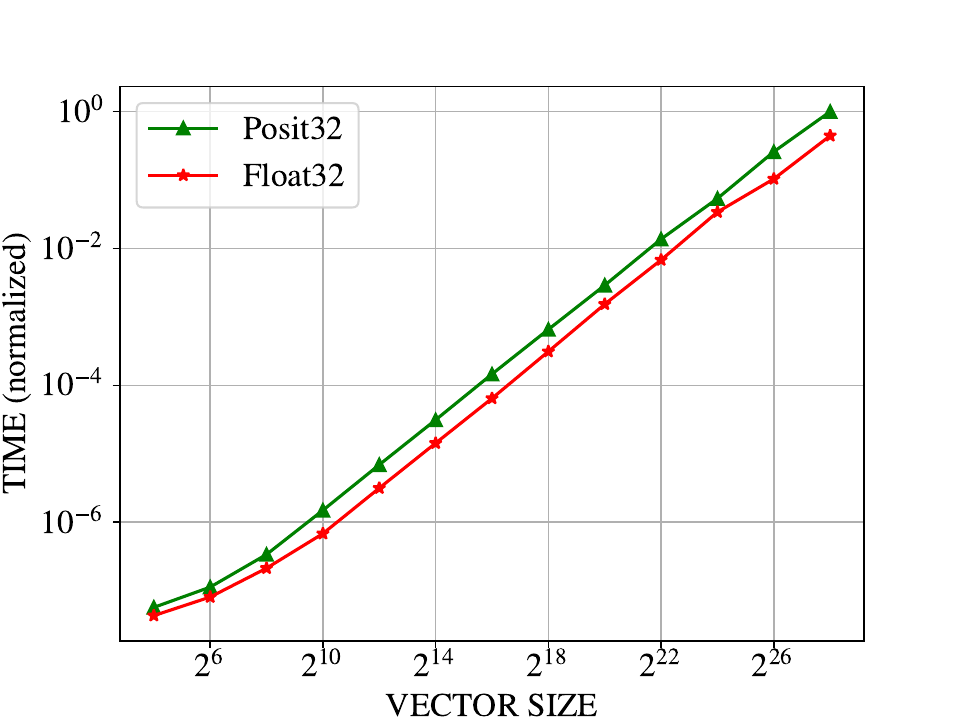}
        \caption{Performance of posit32 vs. float32 on our dataflow architecture.}
        \label{fig:gen1-fft-performance-ieee-vs-p32}
    \end{subfigure}
    \begin{subfigure}[b]{\linewidth}
        \includegraphics[width=\linewidth]{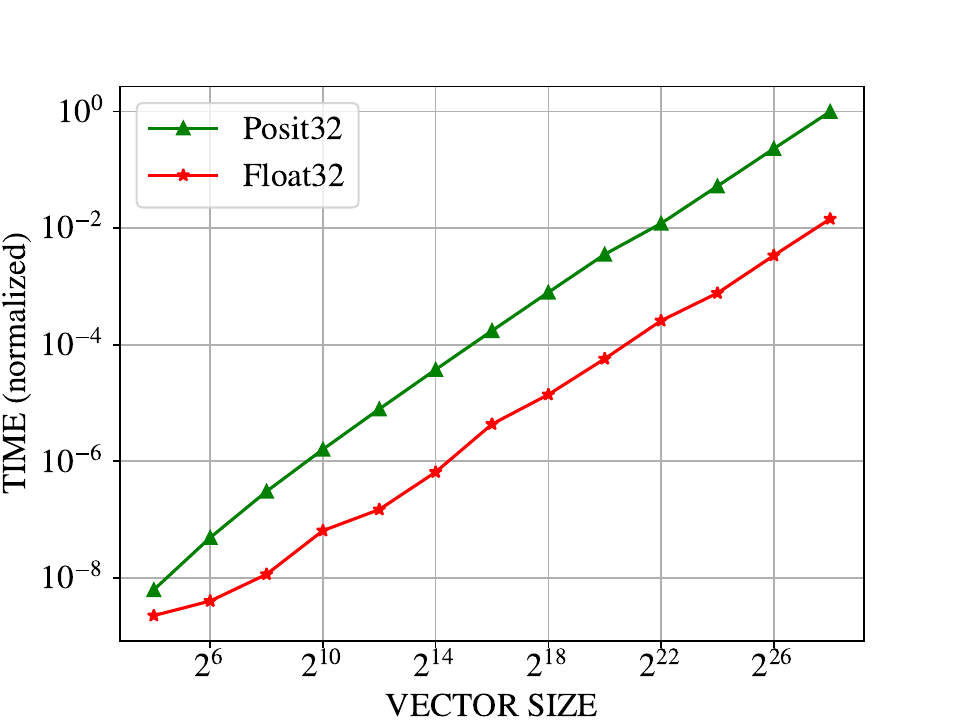}
        \caption{Performance of posit32 vs. float32 on the Intel Xeon 6338.}
        \label{fig:cpu-fft-performance-ieee-vs-p32}
    \end{subfigure}
    \caption{Performance comparison of FFT followed by IFFT for posit32 and float32. The values in graphs are normalized to the highest value in each graph, i.e. the time for posit32 for the $\mathbf{2^{28}}$ case.}
    \label{fig:performance-ieee-vs-p32}
\end{figure}

Fig.~\ref{fig:performance-ieee-vs-p32} shows the normalized time taken to compute a 1D FFT followed by an IFFT for a 1D vector using posit32 and float32. The $x$-axis shows the size of the input data and the $y$-axis shows the normalized time taken for FFT followed by IFFT. We use the same data as shown in Fig.~\ref{fig:hist-inputs} for these experiments. The absolute time is not shown in these measurements to emphasize the performance differences between the dataflow architecture and general-purpose CPUs.

Fig.~\ref{fig:gen1-fft-performance-ieee-vs-p32} shows the normalized performance of posit32 and float32 on our software-defined dataflow architecture, and Fig.~\ref{fig:cpu-fft-performance-ieee-vs-p32} on a Intel Xeon 6338 using 12 physical cores. Fig.~\ref{fig:gen1-fft-performance-ieee-vs-p32} and Fig.~\ref{fig:cpu-fft-performance-ieee-vs-p32} show that the difference in performance of posit32 and float32 is much less when using our dataflow architecture than when using the Intel Xeon CPU. Concretely, Table~\ref{tab:perf-difference-posit-ieee} shows the difference in the performance of posit32 vs. float32 on our dataflow architecture and on the CPU.

Table~\ref{tab:perf-difference-posit-ieee} shows that the CPU performance of posit32 compared to float32 is about $2.77\times$ slower for a vector size of $2^{4}$ and increases to $69.27\times$ for a vector size of $2^{28}$. The CPU implementation is heavily biased in favour of floats as a result of a hardware implementation of normal cases of the format along with other factors which will be explained later. However, the posit implementation for our dataflow architecture is only $1.31\times$ slower for a vector size of $2^{4}$ and $1.82\times$ for a vector size of $2^{28}$. 

\begin{table}[]
\centering
\begin{tabular}{|c|c|c|c}
\cline{1-3}
\textbf{Vector Size} & \textbf{\begin{tabular}[c]{@{}c@{}}Dataflow\\ $\text{Posit}/\text{Float}$\end{tabular}} & \textbf{\begin{tabular}[c]{@{}c@{}}CPU\\ $\text{Posit}/\text{Float}$\end{tabular}} & \textbf{} \\ \cline{1-3}
$\mathbf{2^4}$           & 1.31  & 2.77                                                                           &           \\ \cline{1-3}
$\mathbf{2^6}$           & 1.39  & 12.25                                                                           &           \\ \cline{1-3}
$\mathbf{2^8}$          & 1.58  & 26.47                                                                          &           \\ \cline{1-3}
$\mathbf{2^{10}}$        & 2.19  & 24.81                                                                          &           \\ \cline{1-3}
$\mathbf{2^{12}}$        & 2.19  & 52.94                                                                          &           \\ \cline{1-3}
$\mathbf{2^{14}}$       & 2.18  & 57.41                                                                          &           \\ \cline{1-3}
$\mathbf{2^{16}}$       & 2.26  & 56.11                                                                          &           \\ \cline{1-3}
$\mathbf{2^{18}}$      & 2.10  & 56.77                                                                          &           \\ \cline{1-3}
$\mathbf{2^{20}}$    & 2.11  & 62.00                                                                          &           \\ \cline{1-3}
$\mathbf{2^{22}}$    & 2.01 &  66.67                                                                     &           \\ \cline{1-3}
$\mathbf{2^{24}}$       &   1.83 &   68.41                                                                      &           \\ \cline{1-3}
$\mathbf{2^{26}}$       &    1.83 &     68.16                                                                     &           \\ \cline{1-3}
$\mathbf{2^{28}}$       &    1.82 &     69.27                                                                      &           \\ \cline{1-3}
\end{tabular}
\caption{Difference in performance between posit32 and float32 on our dataflow architecture.}
\label{tab:perf-difference-posit-ieee}
\end{table}

The posit32 FFT is slower than float32 by a narrow margin on the dataflow architecture. An ideal throughput and latency will ensure similar performance for both. However, the hardware imposes some limitations on the performance. Table~\ref{tab:fft-radix-4-ieee-posit} shows that the latency for posit32 is higher than latency for float32. Although the latency is not twice as much higher, the DAG of posit32 is spread across 3 tiles instead of 2 tiles for float32. This causes excessive thread context changes on our architecture, which limits the throughput of the posit32 FFT and IFFT algorithms. Therefore, the reduced performance of posit FFT+IFFT compared to floats is due to a combination of factors such as more operations for basic arithmetic, the resulting longer DAG and its larger latency, and the fact that this longer DAG spans multiple tiles and limits the throughput of the algorithm. A discussion on the impact of latency and throughput can be seen in Sec.~\ref{sec:threading-model}.

\begin{table}[]
\centering
\begin{tabular}{|c|c|c|}
\hline
       & \textbf{Posit32} & \textbf{Float32}  \\ \hline
\textbf{\begin{tabular}[c]{@{}c@{}}Reciprocal throughput \end{tabular}} &   660.5    &       53.25    \\ \hline
\end{tabular}
\caption{Reciprocal throughput of the FFT innermost loop assuming no cache latency for Intel Xeon 6338 from Fig.~\ref{fig:cpu-fft-performance-ieee-vs-p32}.}
\label{tab:posit-ieee-cpu-measurements}
\end{table}

Fig.~\ref{fig:cpu-fft-performance-ieee-vs-p32} shows that the performance of float32 is much faster that of posit32 on the CPU. The difference can be understood by analyzing the throughput on the Ice Lake micro-architecture used by the Intel Xeon 6338. Table~\ref{tab:posit-ieee-cpu-measurements} describes the peak reciprocal throughput of the innermost loop of the radix-4 FFT for float32 and posit32 using uiCA~\cite{Abel19a}. We do not describe IFFT since the radix-4 kernel has an identical number of floating-point operations.

The fact that the posit32 computations have a reciprocal throughput that is more than $10$ times higher than both float32 and float64, coupled with the memory~bound nature of the FFT, explains the difference in the performance of posits compared to floats for the CPU.

\subsubsection{Arithmetic operations of normal float32 and posit32}
\label{sec:arithmetic-operations-ieee754-posit}

As shown in Sec.~\ref{sec:compare-data-flow-with-pipeline}, our software-defined dataflow architecture can express any operation using basic operations as a DAG and execute these operations directly on hardware. The DAG is composed of distinct LEs that are connected with some data dependency. Table~\ref{tab:le-distribution-basic-operators} shows the composition of the DAG for add, subtract and multiply operators for posits compared to floats on our architecture. We do not account for division since it is not used in the algorithms that we test. Each row then represents the type of LEs present for each operator of posit or float arithmetic. The last row shows the total number of LEs for each operator.

It can be seen that all posit32 operations consume more LEs than their float32 counterparts. This can be mainly attributed to the expensive encode and decode operations for posit as shown in Alg.~\ref{alg:p32-decode}. Although the variable length fraction, regime and exponent fields give posits an advantage with regards to relative accuracy, this comes at the additional cost of having to read these fields using more operations than for normal floats. If float32 were made to handle subnormal floats, the performance would be much more similar since decoding then involves the same count-leading-zeros task.

Table~\ref{tab:le-tree-desc-table} reports the height and width of each operation of posit32 and float32. The height of the posit operations is greater than that of the float operations, largely because the regime size must be found before the exponent and fraction can be decoded; floats can decode the scaling factor and significand concurrently. Each LE can show a different latency, and LEs that are independent of each other can be scheduled in parallel using multiple execution units, similar to that in a CPU. Therefore, the total latency of each operation depends on both its height and the LE types it uses.

\begin{table}[]
\centering
\begin{tabular}{|c|ccc|ccc|}
\hline
\multirow{2}{*}{\textbf{}} & \multicolumn{3}{c|}{\textbf{Posit32}}  & \multicolumn{3}{c|}{\textbf{Float32}} \\ \cline{2-7} 
     & \multicolumn{1}{c|}{\textbf{ADD}} & \multicolumn{1}{c|}{\textbf{SUB}} & \textbf{MUL} & \multicolumn{1}{c|}{\textbf{ADD}} & \multicolumn{1}{c|}{\textbf{SUB}} & \textbf{MUL} \\ \hline
\textbf{Height}            & \multicolumn{1}{c|}{90}           & \multicolumn{1}{c|}{92}           & 78           & \multicolumn{1}{c|}{21}   & \multicolumn{1}{c|}{21}  & 12 \\ \hline
\textbf{Width}             & \multicolumn{1}{c|}{5}            & \multicolumn{1}{c|}{5}            & 4            & \multicolumn{1}{c|}{3}    & \multicolumn{1}{c|}{4}   & 3   \\ \hline
\end{tabular}
\vspace{1mm}
\caption{Height and width of basic arithmetic operators (in number of LEs) on our software-defined data flow architecture.}
\label{tab:le-tree-desc-table}
\end{table}

\subsubsection{Hardware projection of the FFT}
\label{sec:projection-on-dfa}

Table~\ref{tab:fft-radix-4-ieee-posit} shows the comparison between the hardware projection of the FFT between posit32 and float32. All results are normalized with respect to posit32 to emphasize the difference between the formats and not the absolute numbers. We do not report numbers for the IFFT since the structure of the DAG is identical to the FFT. The first row indicates the number of logical elements (LEs) that are present in the projection of the entire algorithm. This is analogous to the total LEs shown for the basic arithmetic operators in Sec.~\ref{sec:arithmetic-operations-ieee754-posit}. The LE utilization shows the number of LEs that are actually used out of the total number of LEs in the clusters occupied by the DAG relative to posit32. The area of the projection is normalized according to the number of `clusters' occupied by the LEs, as described in Sec.~\ref{sec:mapping-a-dag-to-hardware}.

\begin{table}[]
\centering
\begin{tabular}{|c|c|c|}
\hline
\textbf{}   & \textbf{Posit32} & \textbf{Float32}  \\ \hline
\textbf{LEs}      & 1  & 0.20 \\ \hline
\textbf{LE utilization} & 1 & 0.76  \\ \hline
\textbf{Latency}  & 1  & 0.64  \\ \hline
\textbf{Area}     & 1  & 0.86\\ \hline
\textbf{Height} & 1 & 0.66   \\ \hline
\end{tabular}
\caption{Comparison of the projection of the 1D FFT for posits and floats on our dataflow architecture. All measurements are normalized according to the posit measurements.}
\label{tab:fft-radix-4-ieee-posit}
\end{table}

A large DAG is split into multiple clusters by our projection algorithm. The latency is reported as the number of clock cycles it would take for a single thread to pass through the entire DAG using the threading model shown in Sec.~\ref{sec:threading-model}.

\subsubsection{Power consumption of the FFT}
\label{sec:power-consumption-dfa}

Table~\ref{tab:power-consumption} shows the power consumption of the FFT followed by the IFFT for a vector size of $262144$. We measured the power consumption in watts and then normalized the values to posit32 watts. The average power consumption is the same for all vector sizes (not shown). It can be seen that our dataflow architecture takes the most power for the posit32 algorithm. This is because of the higher LE utilization of the posit algorithm as shown in Table \ref{tab:fft-radix-4-ieee-posit}. A similar trend can be seen between float32 and float64, where the power consumption of float64 is higher than that of the float32 as a result of higher LE utilization.

\begin{table}[]
\centering
\begin{tabular}{|c|c|c|}
\hline
& \textbf{Posit32} & \textbf{Float32} \\ \hline
\textbf{Power consumption} & 1               & 0.965  \\ \hline
\end{tabular}
\caption{Normalized (average) power usage for the FFT followed by IFFT for a vector size of $\mathbf{2^{18}}$  on our chip. All measurements are normalized according to the posit measurement.}
\label{tab:power-consumption}
\end{table}

\section{Conclusion}
\label{sec:conclufor several appsion}

Posits have been proposed as an alternative to IEEE~754 to overcome some of the limitations of IEEE~754 such as loss of accuracy. We empirically show that posit is more accurate for important algorithms such as spectral analysis and FFT when using the same bit length as IEEE~754. Posit format is most competitive in terms of accuracy when the values are clustered around $0$ in the range $[-1,1]$, achieving about $2\times$ better accuracy than IEEE~754 for the same bit length.

Our implementation of the posit and IEEE~754 (normals only) floating-point specification makes use of a unique software-defined dataflow architecture that implements both specifications using elementary integer operations. Using this architecture, we perform the first comparison using an actual hardware implementation of the posit and IEEE~754 specifications for an important application such as spectral analysis. Our implementation allows us to overcome the limitations of currently available comparisons which either report results obtained via VLSI synthesis or compare software-based posit implementations with hardware-optimized IEEE~754 implementations. Such limitations can be mainly attributed to the cost and time constraints of building hardware from scratch. 

Our implementation allows us to propose a new lower bound for the performance of posits compared to floats. Our 32-bit posit implementation uses about $5\times$ more operations than the 32-bit IEEE~754 floating point format. In spite of this, our posit implementation is only $1.31\times$ slower for a problem size of $2^4$ and $1.82\times$ slower for a problem size of $2^{28}$ on our dataflow architecture. This is in stark contrast to the CPU implementation which is about $2.77\times$ slower for a problem size of $2^4$ and $69.27\times$ for a problem size of $2^{28}$. The CPU implementation is heavily biased in favour of IEEE~754 as a result of the high optimization of IEEE~754 instructions for normal floats. The integer-based software simulation of posit arithmetic on x86 CPUs severly limits its performance.

In the future, we believe our dataflow architecture will allow posits to surpass or match the performance of floats. We plan to do this by optimizing  (via instruction and chain dependency reduction) our posit implementation further (for example the posit encoding and decoding algorithms) in order to reduce the latency of the posit DAG.

\bibliographystyle{ACM-Reference-Format}
\bibliography{Floating_Point,sameer_research_notes}


\begin{thebibliography}{22}


\ifx \showCODEN    \undefined \def \showCODEN     #1{\unskip}     \fi
\ifx \showDOI      \undefined \def \showDOI       #1{#1}\fi
\ifx \showISBNx    \undefined \def \showISBNx     #1{\unskip}     \fi
\ifx \showISBNxiii \undefined \def \showISBNxiii  #1{\unskip}     \fi
\ifx \showISSN     \undefined \def \showISSN      #1{\unskip}     \fi
\ifx \showLCCN     \undefined \def \showLCCN      #1{\unskip}     \fi
\ifx \shownote     \undefined \def \shownote      #1{#1}          \fi
\ifx \showarticletitle \undefined \def \showarticletitle #1{#1}   \fi
\ifx \showURL      \undefined \def \showURL       {\relax}        \fi
\providecommand\bibfield[2]{#2}
\providecommand\bibinfo[2]{#2}
\providecommand\natexlab[1]{#1}
\providecommand\showeprint[2][]{arXiv:#2}

\bibitem[noa({[n.\,d.]})]%
        {noauthor_ieee_2019}
 \bibinfo{year}{[n.\,d.]}\natexlab{}.
\newblock \showarticletitle{{IEEE} Standard for Floating-Point Arithmetic}.
\newblock  (\bibinfo{year}{[n.\,d.]}), \bibinfo{pages}{1--84}.
\newblock
\showISSN{null}
\urldef\tempurl%
\url{https://doi.org/10.1109/IEEESTD.2019.8766229}
\showDOI{\tempurl}


\bibitem[Abel and Reineke(2019)]%
        {Abel19a}
\bibfield{author}{\bibinfo{person}{Andreas Abel} {and} \bibinfo{person}{Jan Reineke}.} \bibinfo{year}{2019}\natexlab{}.
\newblock \showarticletitle{uops.info: Characterizing Latency, Throughput, and Port Usage of Instructions on Intel Microarchitectures}. In \bibinfo{booktitle}{\emph{ASPLOS}} (Providence, RI, USA) \emph{(\bibinfo{series}{ASPLOS '19})}. \bibinfo{publisher}{ACM}, \bibinfo{address}{New York, NY, USA}, \bibinfo{pages}{673--686}.
\newblock
\showISBNx{978-1-4503-6240-5}
\urldef\tempurl%
\url{https://doi.org/10.1145/3297858.3304062}
\showDOI{\tempurl}


\bibitem[Bronstein et~al\mbox{.}(2017)]%
        {Bronstein_2017}
\bibfield{author}{\bibinfo{person}{Michael~M. Bronstein}, \bibinfo{person}{Joan Bruna}, \bibinfo{person}{Yann LeCun}, \bibinfo{person}{Arthur Szlam}, {and} \bibinfo{person}{Pierre Vandergheynst}.} \bibinfo{year}{2017}\natexlab{}.
\newblock \showarticletitle{Geometric Deep Learning: Going beyond Euclidean data}.
\newblock \bibinfo{journal}{\emph{{IEEE} Signal Processing Magazine}} \bibinfo{volume}{34}, \bibinfo{number}{4} (\bibinfo{date}{jul} \bibinfo{year}{2017}), \bibinfo{pages}{18--42}.
\newblock
\urldef\tempurl%
\url{https://doi.org/10.1109/msp.2017.2693418}
\showDOI{\tempurl}


\bibitem[Bruna et~al\mbox{.}(2014)]%
        {bruna2014spectral}
\bibfield{author}{\bibinfo{person}{Joan Bruna}, \bibinfo{person}{Wojciech Zaremba}, \bibinfo{person}{Arthur Szlam}, {and} \bibinfo{person}{Yann LeCun}.} \bibinfo{year}{2014}\natexlab{}.
\newblock \bibinfo{title}{Spectral Networks and Locally Connected Networks on Graphs}.
\newblock
\newblock
\showeprint[arxiv]{1312.6203}~[cs.LG]


\bibitem[Carmichael et~al\mbox{.}(2019)]%
        {carmichael2019}
\bibfield{author}{\bibinfo{person}{Zachariah Carmichael}, \bibinfo{person}{Hamed~F. Langroudi}, \bibinfo{person}{Char Khazanov}, \bibinfo{person}{Jeffrey Lillie}, \bibinfo{person}{John~L. Gustafson}, {and} \bibinfo{person}{Dhireesha Kudithipudi}.} \bibinfo{year}{2019}\natexlab{}.
\newblock \bibinfo{title}{Deep {{Positron}}: {{A Deep Neural Network Using}} the {{Posit Number System}}}.
\newblock
\newblock
\showeprint[arxiv]{1812.01762}~[cs]


\bibitem[Chien et~al\mbox{.}(2020)]%
        {chien2020}
\bibfield{author}{\bibinfo{person}{Steven W.~D. Chien}, \bibinfo{person}{Ivy~B. Peng}, {and} \bibinfo{person}{Stefano Markidis}.} \bibinfo{year}{2020}\natexlab{}.
\newblock \showarticletitle{Posit {{NPB}}: {{Assessing}} the {{Precision Improvement}} in {{HPC Scientific Applications}}}.
\newblock In \bibinfo{booktitle}{\emph{Parallel {{Processing}} and {{Applied Mathematics}}}}, \bibfield{editor}{\bibinfo{person}{Roman Wyrzykowski}, \bibinfo{person}{Ewa Deelman}, \bibinfo{person}{Jack Dongarra}, {and} \bibinfo{person}{Konrad Karczewski}} (Eds.). Vol.~\bibinfo{volume}{12043}. \bibinfo{publisher}{{Springer International Publishing}}, \bibinfo{address}{{Cham}}, \bibinfo{pages}{301--310}.
\newblock
\showISBNx{978-3-030-43228-7 978-3-030-43229-4}
\urldef\tempurl%
\url{https://doi.org/10.1007/978-3-030-43229-4_26}
\showDOI{\tempurl}


\bibitem[Ciocirlan et~al\mbox{.}(2021)]%
        {ciocirlan2021}
\bibfield{author}{\bibinfo{person}{Stefan~Dan Ciocirlan}, \bibinfo{person}{Dumitrel Loghin}, \bibinfo{person}{Lavanya Ramapantulu}, \bibinfo{person}{Nicolae Ţăpuş}, {and} \bibinfo{person}{Yong~Meng Teo}.} \bibinfo{year}{2021}\natexlab{}.
\newblock \showarticletitle{The Accuracy and Efficiency of Posit Arithmetic}. In \bibinfo{booktitle}{\emph{2021 IEEE 39th International Conference on Computer Design (ICCD)}}. \bibinfo{pages}{83--87}.
\newblock
\urldef\tempurl%
\url{https://doi.org/10.1109/ICCD53106.2021.00024}
\showDOI{\tempurl}


\bibitem[De~Silva et~al\mbox{.}(2023)]%
        {Silva2023}
\bibfield{author}{\bibinfo{person}{Himeshi De~Silva}, \bibinfo{person}{Hongshi Tan}, \bibinfo{person}{Nhut-Minh Ho}, \bibinfo{person}{John~L. Gustafson}, {and} \bibinfo{person}{Weng-Fai Wong}.} \bibinfo{year}{2023}\natexlab{}.
\newblock \showarticletitle{Towards A Better 16-Bit Number Representation For Training Neural Networks}. In \bibinfo{booktitle}{\emph{Next Generation Arithmetic: 4th International Conference, CoNGA 2023, Singapore, March 1-2, 2023, Proceedings}} (Singapore, Singapore). \bibinfo{publisher}{Springer-Verlag}, \bibinfo{address}{Berlin, Heidelberg}, \bibinfo{pages}{114–133}.
\newblock
\showISBNx{978-3-031-32179-5}
\urldef\tempurl%
\url{https://doi.org/10.1007/978-3-031-32180-1_8}
\showDOI{\tempurl}


\bibitem[Fousse et~al\mbox{.}(2007)]%
        {Fousse2007}
\bibfield{author}{\bibinfo{person}{Laurent Fousse}, \bibinfo{person}{Guillaume Hanrot}, \bibinfo{person}{Vincent Lef\`{e}vre}, \bibinfo{person}{Patrick P\'{e}lissier}, {and} \bibinfo{person}{Paul Zimmermann}.} \bibinfo{year}{2007}\natexlab{}.
\newblock \showarticletitle{MPFR: A Multiple-Precision Binary Floating-Point Library with Correct Rounding}.
\newblock \bibinfo{journal}{\emph{ACM Trans. Math. Softw.}} \bibinfo{volume}{33}, \bibinfo{number}{2} (\bibinfo{date}{June} \bibinfo{year}{2007}), \bibinfo{pages}{13–es}.
\newblock
\showISSN{0098-3500}
\urldef\tempurl%
\url{https://doi.org/10.1145/1236463.1236468}
\showDOI{\tempurl}


\bibitem[Franchetti et~al\mbox{.}(2018)]%
        {franchetti2018}
\bibfield{author}{\bibinfo{person}{Franz Franchetti}, \bibinfo{person}{Daniele~G. Spampinato}, \bibinfo{person}{Anuva Kulkarni}, \bibinfo{person}{Doru Thom~Popovici}, \bibinfo{person}{Tze~Meng Low}, \bibinfo{person}{Michael Franusich}, \bibinfo{person}{Andrew Canning}, \bibinfo{person}{Peter McCorquodale}, \bibinfo{person}{Brian~Van Straalen}, {and} \bibinfo{person}{Phillip Colella}.} \bibinfo{year}{2018}\natexlab{}.
\newblock \showarticletitle{{{FFTX}} and {{SpectralPack}}: {{A First Look}}}. In \bibinfo{booktitle}{\emph{2018 {{IEEE}} 25th {{International Conference}} on {{High Performance Computing Workshops}} ({{HiPCW}})}}. \bibinfo{publisher}{{IEEE}}, \bibinfo{address}{{Bengaluru, India}}, \bibinfo{pages}{18--27}.
\newblock
\showISBNx{978-1-72810-114-9}
\urldef\tempurl%
\url{https://doi.org/10.1109/HiPCW.2018.8634111}
\showDOI{\tempurl}


\bibitem[Goldberg(1991)]%
        {goldberg1991}
\bibfield{author}{\bibinfo{person}{David Goldberg}.} \bibinfo{year}{1991}\natexlab{}.
\newblock \showarticletitle{What Every Computer Scientist Should Know about Floating-Point Arithmetic}.
\newblock \bibinfo{journal}{\emph{Comput. Surveys}} \bibinfo{volume}{23}, \bibinfo{number}{1} (\bibinfo{date}{March} \bibinfo{year}{1991}), \bibinfo{pages}{5--48}.
\newblock
\showISSN{0360-0300}
\urldef\tempurl%
\url{https://doi.org/10.1145/103162.103163}
\showDOI{\tempurl}


\bibitem[Gustafson and Yonemoto({[n.\,d.]})]%
        {gustafson_beating_2017}
\bibfield{author}{\bibinfo{person}{Gustafson} {and} \bibinfo{person}{Yonemoto}.} \bibinfo{year}{[n.\,d.]}\natexlab{}.
\newblock \showarticletitle{Beating Floating Point at its Own Game}.
\newblock  (\bibinfo{year}{[n.\,d.]}).
\newblock
\urldef\tempurl%
\url{https://dl.acm.org/doi/abs/10.14529/jsfi170206}
\showURL{%
\tempurl}


\bibitem[Gustafson(2022)]%
        {gustafson2022}
\bibfield{author}{\bibinfo{person}{John~L Gustafson}.} \bibinfo{year}{2022}\natexlab{}.
\newblock \showarticletitle{A {{Generalized Framework}} for {{Matching Arithmetic Format}} to {{Application Requirements}}}.
\newblock  (\bibinfo{year}{2022}), \bibinfo{pages}{9}.
\newblock


\bibitem[Leong and Gustafson(2023)]%
        {leong2023}
\bibfield{author}{\bibinfo{person}{Siew~Hoon Leong} {and} \bibinfo{person}{John~L. Gustafson}.} \bibinfo{year}{2023}\natexlab{}.
\newblock \showarticletitle{Lossless {{FFTs Using Posit Arithmetic}}}. In \bibinfo{booktitle}{\emph{Next {{Generation Arithmetic}}}} \emph{(\bibinfo{series}{Lecture {{Notes}} in {{Computer Science}}})}, \bibfield{editor}{\bibinfo{person}{John Gustafson}, \bibinfo{person}{Siew~Hoon Leong}, {and} \bibinfo{person}{Marek Michalewicz}} (Eds.). \bibinfo{publisher}{{Springer Nature Switzerland}}, \bibinfo{address}{{Cham}}, \bibinfo{pages}{1--18}.
\newblock
\showISBNx{978-3-031-32180-1}
\urldef\tempurl%
\url{https://doi.org/10.1007/978-3-031-32180-1_1}
\showDOI{\tempurl}


\bibitem[Montero et~al\mbox{.}(2019)]%
        {montero2019}
\bibfield{author}{\bibinfo{person}{Ra{\'u}l~Murillo Montero}, \bibinfo{person}{Alberto~A. Del~Barrio}, {and} \bibinfo{person}{Guillermo Botella}.} \bibinfo{year}{2019}\natexlab{}.
\newblock \bibinfo{title}{Template-{{Based Posit Multiplication}} for {{Training}} and {{Inferring}} in {{Neural Networks}}}.
\newblock
\newblock
\showeprint[arxiv]{1907.04091}~[cs]


\bibitem[Murillo et~al\mbox{.}({[n.\,d.]})]%
        {murillo_comparing_2022}
\bibfield{author}{\bibinfo{person}{Raul Murillo}, \bibinfo{person}{David Mallasén}, \bibinfo{person}{Alberto~A. Del~Barrio}, {and} \bibinfo{person}{Guillermo Botella}.} \bibinfo{year}{[n.\,d.]}\natexlab{}.
\newblock \showarticletitle{Comparing Different Decodings for Posit Arithmetic}. In \bibinfo{booktitle}{\emph{Next Generation Arithmetic}} (Cham, 2022) \emph{(\bibinfo{series}{Lecture Notes in Computer Science})}, \bibfield{editor}{\bibinfo{person}{John Gustafson} {and} \bibinfo{person}{Vassil Dimitrov}} (Eds.). \bibinfo{publisher}{Springer International Publishing}, \bibinfo{pages}{84--99}.
\newblock
\showISBNx{978-3-031-09779-9}
\urldef\tempurl%
\url{https://doi.org/10.1007/978-3-031-09779-9_6}
\showDOI{\tempurl}


\bibitem[Ootomo et~al\mbox{.}(2023)]%
        {ootomo2023}
\bibfield{author}{\bibinfo{person}{Hiryuki Ootomo}, \bibinfo{person}{Hidetaka Manabe}, \bibinfo{person}{Kenji Harada}, {and} \bibinfo{person}{Rio Yokota}.} \bibinfo{year}{2023}\natexlab{}.
\newblock \bibinfo{title}{Quantum {{Circuit Simulation}} by {{SGEMM Emulation}} on {{Tensor Cores}} and {{Automatic Precision Selection}}}.
\newblock
\newblock
\showeprint[arxiv]{2303.08989}~[quant-ph]


\bibitem[Ozaki et~al\mbox{.}(2012)]%
        {ozaki2012}
\bibfield{author}{\bibinfo{person}{Katsuhisa Ozaki}, \bibinfo{person}{Takeshi Ogita}, \bibinfo{person}{Shin'ichi Oishi}, {and} \bibinfo{person}{Siegfried~M. Rump}.} \bibinfo{year}{2012}\natexlab{}.
\newblock \showarticletitle{Error-Free Transformations of Matrix Multiplication by Using Fast Routines of Matrix Multiplication and Its Applications}.
\newblock \bibinfo{journal}{\emph{Numerical Algorithms}} \bibinfo{volume}{59}, \bibinfo{number}{1} (\bibinfo{date}{Jan.} \bibinfo{year}{2012}), \bibinfo{pages}{95--118}.
\newblock
\showISSN{1572-9265}
\urldef\tempurl%
\url{https://doi.org/10.1007/s11075-011-9478-1}
\showDOI{\tempurl}


\bibitem[Palacz(2018)]%
        {Palacz2018}
\bibfield{author}{\bibinfo{person}{Magdalena Palacz}.} \bibinfo{year}{2018}\natexlab{}.
\newblock \showarticletitle{Spectral Methods for Modelling of Wave Propagation in Structures in Terms of Damage Detection—A Review}.
\newblock \bibinfo{journal}{\emph{Applied Sciences}} \bibinfo{volume}{8}, \bibinfo{number}{7} (\bibinfo{year}{2018}).
\newblock
\showISSN{2076-3417}
\urldef\tempurl%
\url{https://doi.org/10.3390/app8071124}
\showDOI{\tempurl}


\bibitem[Parikh et~al\mbox{.}(2023)]%
        {parikh2023}
\bibfield{author}{\bibinfo{person}{Devangi~N. Parikh}, \bibinfo{person}{Robert~A. Geijn}, {and} \bibinfo{person}{Greg~M. Henry}.} \bibinfo{year}{2023}\natexlab{}.
\newblock \showarticletitle{Cascading GEMM: High Precision from Low Precision}.
\newblock \bibinfo{journal}{\emph{ArXiv}}  \bibinfo{volume}{abs/2303.04353} (\bibinfo{year}{2023}).
\newblock
\urldef\tempurl%
\url{https://api.semanticscholar.org/CorpusID:257405236}
\showURL{%
\tempurl}


\bibitem[Sharma et~al\mbox{.}(2021)]%
        {sharma2021}
\bibfield{author}{\bibinfo{person}{Niraj Sharma}, \bibinfo{person}{Riya Jain}, \bibinfo{person}{Madhumita Mohan}, \bibinfo{person}{Sachin Patkar}, \bibinfo{person}{Rainer Leupers}, \bibinfo{person}{Nikhil Rishiyur}, {and} \bibinfo{person}{Farhad Merchant}.} \bibinfo{year}{2021}\natexlab{}.
\newblock \bibinfo{title}{{{CLARINET}}: {{A RISC-V Based Framework}} for {{Posit Arithmetic Empiricism}}}.
\newblock
\newblock
\showeprint[arxiv]{2006.00364}~[cs]


\bibitem[Trefethen(2000)]%
        {trefethen2000}
\bibfield{author}{\bibinfo{person}{Lloyd~N. Trefethen}.} \bibinfo{year}{2000}\natexlab{}.
\newblock \bibinfo{booktitle}{\emph{Spectral {{Methods}} in {{MATLAB}}}}.
\newblock \bibinfo{publisher}{{Society for Industrial and Applied Mathematics}}.
\newblock
\showISBNx{978-0-89871-465-4}
\urldef\tempurl%
\url{https://doi.org/10.1137/1.9780898719598}
\showDOI{\tempurl}


\end{thebibliography}

\end{document}